\documentclass[11pt]{article}
\usepackage{amsfonts}
\usepackage{amssymb}
\usepackage{amsthm}
\usepackage{epsf}

\def\GG{{\rm I}\!\Gamma}
\def\G{\Gamma}

\begin{document}

\begin{flushright}
{\bf IFUM 654/FT} \\
\end{flushright}
\vskip 2 truecm
{
\Large
\bf
\centerline{Higher Order Anomaly Consistency Conditions:}
\centerline{Renormalization and Non-Locality}
\normalsize
\rm
\vspace{0.6cm} 
}

\begin{center}
{\Large Marco Picariello \footnote{marco.picariello@mi.infn.it} and 
 Andrea Quadri \footnote{andrea.quadri@mi.infn.it}}
\\
\vskip 0.5 truecm 
Dipartimento di Fisica, Universit\`a di Milano
via Celoria 16, 20133 Milano, Italy \\
and  INFN, Sezione di Milano \\

\end{center}

\setlength {\baselineskip}{0.2in}
\vskip 0.6 cm
\normalsize
\bf

\vskip 1 cm
\centerline{Abstract}
\rm
\small
\begin{quotation}
We study the Slavnov-Taylor Identities (STI) breaking terms, 
up to the second order in perturbation theory.
\\
We investigate which requirements are needed for the first order
Wess-Zumino consistency condition to hold true at the next order in
perturbation theory. We find that:
\begin{description}
\item[a)]{If 
the cohomologically trivial contributions to the first order ST breaking terms
are not removed by a suitable choice of the first order
normalization conditions,
the Wess-Zumino consistency condition is modified
at the second order.}
\item[b)]{
Moreover, if one fails to recover
the cohomologically trivial part of the first order STI breaking local
functional, 
the second order anomaly actually turns out to be a non-local
functional of the fields and the external sources. 
}
\end{description}
By using the  
Wess-Zumino consistency condition and the Quantum Action Principle,
we show that the cohomological analysis of the first order
STI breaking terms can actually be performed also in
a model (the massive Abelian Higgs-Kibble one)
where the BRST transformations are not nilpotent.
\end{quotation}


\vfill\eject
\section{Introduction}

It is well known that classical symmetries, expressed in
functional form by a set of Ward identities (WI) satisfied by the
classical action $\G^{(0)}$, may be broken at the quantum level
\cite{PS}.
The possible quantum breaking in perturbative QFT of the WI
can have two origins.
Either it is a truly physical obstruction
to the restoration of the WI 
(and one is then faced with the problem
of classifying all possible anomalous terms for the model
under investigation) or it is an
unwanted effect of the renormalization procedure, needed to handle
UV divergences in the full quantum action $\G$.
The latter breaking can always be recovered by the
choice of a more suited renormalization scheme, or of different
renormalization conditions. Since changing the renormalization
procedure does not alter the physical content of the theory, 
this is a spurious breaking of the WI.~\footnote{Though 
non-physical, these breaking terms nevertheless
require a careful treatment. For example, in the Standard Model
no regularization scheme is known to preserve all the symmetries
of the theory, because of the presence of the $\gamma_5$ matrix
and of the completely antisymmetric tensor $\epsilon_{\mu\nu\rho\sigma}$.
Thus some procedures are needed to recover the Ward identities
broken by the intermediate renormalization: suitably defined finite
counter-terms must be added to the regularized quantum action
in order to recover the relevant Ward identities.}

In gauge theories,   
the BRST \cite{BRST} transformations turn out to be a very powerful tool for
proving the unitarity of the physical $S$ matrix \cite{UNIT}. 
Thus the restoration of BRST symmetry is an essential step
in the perturbative construction of gauge theories. 
The crucial requirement
in the proof of physical unitarity is nilpotency: nilpotency of the
BRST transformations
is a sufficient condition for a theory to be unitary, provided that
the associated WI (known as Slavnov-Taylor identities -
STI from now on) hold at the quantum level.
The task is then to restore the STI,
when this is physically possible (absence of anomalies), and to
completely classify the form of the breaking terms in the
anomalous case.

The nilpotency of the BRST transformations also allows for an effective
cohomological analysis of the ST breaking terms.
This is most easily seen in the framework of the field-antifield
formalism, an extension of the original BRST formulation \cite{GPS}.

We consider a general gauge theory with fields $\phi_i$ and ghosts
$c_k$, introduced by the covariant quantization of the model.
The fields $\phi_i$
and the ghosts $c_k$ are collectively denoted by $\Phi^A$,
$A=1, \dots, N$. 
In the field-antifield formalism, for each field $\Phi^A$ one
introduces an antifield $\Phi^*_A$. 
The space ${\cal F}$ of the functionals of $\Phi^A, \Phi^*_A$ is endowed 
with an odd symplectic structure $(\cdot,\cdot)$, the antibracket:
%
\begin{eqnarray}
(X,Y) = \sum_{A=1}^N
\int d^4x \, \left ( 
{\delta_r X \over \delta \Phi^A(x)}
{\delta_l Y \over \delta \Phi_A^*(x)}-
{\delta_r X \over \delta \Phi_A^*(x)}
{\delta_l Y \over \delta \Phi^A(x)}
 \right )\ .
\label{intro1}
\end{eqnarray}
The subscripts $r$ and $l$ denote right and left differentiation
respectively. The classical action $\G^{(0)}[\Phi,\Phi^*]$
is assumed to satisfy the classical master equation
\begin{eqnarray}
(\G^{(0)},\G^{(0)})=0\ ,
\label{intro2}
\end{eqnarray}
under the condition that $\left . \G^{(0)}[\Phi,\Phi^*] \right |_{\Phi^*=0}$
coincides with the classical gauge-fixed BRST invariant action.
The quantization of the theory produces an effective
action $\G$
\begin{eqnarray}
\G = \G^{(0)}+\sum_{n \geq 1} \hbar^n \G^{(n)}\ ,
\label{intro3}
\end{eqnarray}
which satisfies the quantum extension of the classical
master equation (\ref{intro2}) \cite{zj}:
\begin{eqnarray}
{1 \over 2} (\G,\G) = \hbar ({\cal A}\cdot \G)\ ,
\label{intro4}
\end{eqnarray}
where the insertion $({\cal A}\cdot \G)$ represents the possible
anomalous terms due to the quantum corrections.
Using the graded Jacobi identity for the antibracket $(\cdot,\cdot)$
\begin{eqnarray}
((X,Y),Z) + (-1)^{(\epsilon_X+1)(\epsilon_Y+\epsilon_Z)}((Y,Z),X)
+(-1)^{(\epsilon_Z+1)(\epsilon_X+\epsilon_Y)}((Z,X),Y)=0
\label{intro5}
\end{eqnarray}
(where $\epsilon_X=0$ if $X$ obeys Bose statistics and 
$\epsilon_X=1$ if $X$ obeys Fermi statistics), it is easy to deduce
the following identity for $\G$:
\begin{eqnarray}
(\G,(\G,\G))=0
\label{intro6}
\end{eqnarray}
or, taking into account eq.(\ref{intro4}),
\begin{eqnarray}
\hbar (\G,({\cal A}\cdot \G))=0\ .
\label{intro7} 
\end{eqnarray}
At the first order in perturbation theory the previous condition
reduces to
\begin{eqnarray}
(\G^{(0)},{\cal A}_1 )=0\ ,
\label{intro8}
\end{eqnarray}
where ${\cal A} \cdot \G = \sum_{j \geq 1} \hbar^{j-1} {\cal A}_j$.
Eq.(\ref{intro8}) is the Wess-Zumino consistency condition \cite{wz} written
in the field-antifield formalism.

Since $\G^{(0)}$ satisfies the classical master equation (\ref{intro2}),
the operator $(\G^{(0)},\cdot)$ is nilpotent and eq.(\ref{intro8}) gives rise
to a cohomological problem. The most general solution of eq.(\ref{intro8})
can be cast in the form
\begin{eqnarray}
{\cal A}_1 = \sum_{k \geq 1} \lambda_k {\cal C}_k + (\G^{(0)},
{\cal C}_0)\ ,
\label{intro9}
\end{eqnarray}
where the sum is over the representatives ${\cal C}_k$ of the independent
non-trivial cohomology classes of the operator $(\G^{(0)},\cdot)$
and $(\G^{(0)},{\cal C}_0)$ is an arbitrary element of the trivial
cohomology class with FP-charge $1$.
The purely algebraic analysis which leads to eq.(\ref{intro9}) puts no
restrictions on the form of ${\cal C}_k$ and ${\cal C}_0$, apart from
saying that they are functionals of $\Phi^A, \Phi_A^*$. In particular,
${\cal C}_k$ and ${\cal C}_0$ might very well be non-local functionals
of $\Phi^A, \Phi_A^*$ and they might contain arbitrarily high powers
of $\Phi^A,\Phi_A^*$, when expanded on a basis of the space ${\cal F}$.

However, if the theory is power-counting renormalizable and the quantization
is performed by means of a renormalization procedure which satisfies the 
Quantum Action Principle (QAP)~ \cite{PS,QAP}, several strong
restrictions are imposed
on ${\cal C}_k$ and  ${\cal C}_0$: they must be {\em local}
functionals of $\Phi^A$, $\Phi^*_A$ and have dimensions that cannot exceed
a finite upper limit, predicted by the QAP.

Taking into account these constraints on locality and power counting,
one sees from eq.(\ref{intro4}) that $(\G^{(0)},{\cal C}_0)$ can
always be reabsorbed by adding
finite counter-terms to $\G^{(1)}$.  Thus $(\G^{(0)},{\cal C}_0)$  is
a spurious contribution to the anomaly. 

If some of the coefficients $\lambda_k$ actually turn out
to be non-zero (once their calculation has been performed in the 
intermediate renormalization scheme), the theory is truly anomalous: 
no matter how one changes the finite part of $\G^{(1)}$, 
these terms cannot be reabsorbed. Moreover, it is believed
that in every admissible renormalization scheme (compatible with
Poincar\'e invariance and all other exact symmetries of the
theory~\footnote{For the sake of definiteness we are supposing that
the ST invariance is the only broken invariance of the model.})
the calculation of $\lambda^{(n)}$ will yield a non-zero result.

Some attempts have been made to push forward this kind of analysis
of the anomalous terms to higher orders in perturbation theory \cite{white,
paris}.
The strategy is to find out a suitable higher-order generalization
of the Wess-Zumino consistency condition in eq.(\ref{intro8}),
relying on the consistency condition for the full quantum action
$\G$ in eq.(\ref{intro6}) and 
the nilpotency of the operator $(\G^{(0)},\cdot)$.

In this program, the properties of the renormalization procedure and
the choice of the renormalization conditions turn out to be
as crucial as the algebraic features of the cohomological analysis, dictated
by the nilpotency of $(\G^{(0)},\cdot)$. 
 
%
%

%
%
%


In this paper we discuss the second order generalization of the Wess-Zumino
consistency condition,
in the simple framework of 
the Abelian Higgs-Kibble model
\cite{becHK}.
However, the conclusions one can draw from this example are general
enough to be of interest for a wide class of gauge theories.

Although the Abelian HK model exhibits spurious anomalous terms only
(see \cite{FG,FGQ} for a detailed analysis), we prove that,
if the cohomologically trivial  contributions to ${\cal A}_1$
are not recovered by suitably chosen finite counter-terms in
$\G^{(1)}$, at the next
order the equation for the anomaly is no more the Wess-Zumino
consistency condition
\begin{eqnarray}
(\G^{(0)},{\cal A}_2)=0 \ .
\label{k3bis}
\end{eqnarray} 
Moreover, we show that
in this case ${\cal A}_2$ must be non-local, in sharp contrast with
the locality of the solutions of eq.(\ref{k3bis}).
 
In our discussion 
we will relax the assumption of nilpotency of the BRST
transformations, by adding to the classical action the following
mass term
\begin{eqnarray}
\int d^4x \, 
\left ( {M^2 \over 2} A_\mu^2 + M^2 \bar c c - {M^2 \over 2 \alpha} (\phi_1^2 + \phi_2^2)
\right ).
\label{e6}
\end{eqnarray}
Even though the price of this generalization is the loss of unitarity,
in the massive framework it is simpler to appreciate the interplay
between algebraic properties and the behavior of the quantum theory
under the renormalization procedure.
This also allows to discuss the conditions
under which strict nilpotency of $(\G^{(0)},\cdot)$ is actually needed
to carry out the construction of ${\cal A} \cdot \G$, to higher
orders in perturbation theory.

\section{Consistency conditions in the non-nilpotent case}

In the Abelian Higgs-Kibble model the fields $A_\mu$ and $\bar c$
have linear BRST transformations. Thus one can actually 
avoid to introduce their antifields.
Moreover, we work in the on-shell formalism, i.e. we have eliminated
the auxiliary Nakanishi-Lautrup field $B$ \cite{nakanishi}
associated with the BRST variation of $\bar c$ in the off-shell formalism.
This in turn allows for a simplification of the Feynman graphs involved
in our analysis.

From now on we use the reduced antibracket
\begin{eqnarray}
\left(X,Y \right) = 
\int d^4x \left [ {{\delta X}\over{\delta J_1}}{{\delta
Y}\over{\delta\phi_1}} 
+ {{\delta X}\over{\delta J_2}}{{\delta
Y}\over{\delta\phi_2}}  
-{{\delta X}\over{\delta \psi}}{{\delta
Y}\over{\delta{\bar\eta}}} 
+{{\delta X}\over{\delta {\bar\psi}}}{{\delta
Y}\over{\delta\eta}}
\right ] \, .
\label{e1bis}
\end{eqnarray}
All functional derivatives are assumed to act
from the left.
$J_1,J_2,\eta,\bar \eta$ are the antifields of $\phi_1,\phi_2,
\bar \psi, \psi$ respectively.

The ST identities for the Abelian Higgs-Kibble model then read 
\begin{eqnarray}
S(\G)=0 \, ,
\label{STid}
\end{eqnarray}
where the ST operator is
\begin{eqnarray}
S(\G)= \int d^4 x \left[\partial^\mu c \, {\delta \G
\over \delta {A^\mu}}
+\left ( \partial A + {ev \over \alpha} \phi_2\right )
{\delta \G \over \delta {\bar c}} \right] + \left(\G,\G\right) \, .
\label{e1}
\end{eqnarray}

The complete antibracket ${1\over2} (\G,\G)$ in eq.(\ref{intro4}) becomes
the ST operator in eq.(\ref{e1}).

In the on-shell formalism the ghost
equation is:
\begin{eqnarray}
{\cal G} \G = \alpha \square c +M^2 c \, ,
\label{e8}
\end{eqnarray}
where the ghost operator ${\cal G}$ is defined by:
\begin{eqnarray}
{\cal G} = {\delta (\cdot) \over \delta \bar c} - ev {\delta (\cdot) \over \delta J_2} \, .
\label{e14septies}
\end{eqnarray}

\subsection{The massless case}

In the massless (nilpotent) case, eq. (\ref{intro6}) is translated into
\begin{eqnarray}
S_\G (S(\G))=0\ .
\label{e2}
\end{eqnarray}
$S_\G$ denotes the linearization of the ST operator (\ref{e1}):
\begin{eqnarray}
S_\G(\cdot)= \int d^4 x \left[\partial^\mu c \, {\delta (\cdot)
\over \delta {A^\mu}}
+\left ( \partial A + {ev \over \alpha} \phi_2 \right )
{\delta (\cdot) \over \delta {\bar c}} \right] + \left(\G,\cdot\right)
+ \left(\cdot,\G\right) \ .
\label{e3}
\end{eqnarray}
The identity (\ref{e2}) is valid for {\em any} $\G$ (even for
a $\G$ which does not satisfy the STI $S(\G)=0$).

We denote by $S_0 \equiv S_{\G^{(0)}}$ the zero-th order ST linearization.
Notice that 
\begin{eqnarray}
\left \{ {\cal G}, S_0 \right \} = 0 \, .
\label{e14octies}
\end{eqnarray}

We perform a formal expansion for $S(\G)$ in powers of $\hbar$:
\begin{eqnarray}
S(\G) = \sum_{n \geq 0} \hbar^n S(\G)^{(n)}\ .
\label{anomexp}
\end{eqnarray}

At the first order in perturbation theory eq.(\ref{e2}) becomes
\begin{eqnarray}
S_0(S(\G)^{(1)})=0\ ,
\label{e4}
\end{eqnarray}
which parallels eq.(\ref{intro8}) and gives rise to the cohomological
analysis of $S(\G)^{(1)}$. 
Thanks to the nilpotency of $S_0$ (guaranteed by the invariance of the classical action
$S(\G^{(0)}) = 0)$, it is possible to find the most general form of 
$S(\G)^{(1)}$ compatible with 
condition (\ref{e4}). $S(\G)^{(1)}$ can be cast in the form
\begin{eqnarray}
S(\G)^{(1)} = Y^{(1)} + S_0({\cal C}_0)  \, ,
\label{e5}
\end{eqnarray}
where $Y^{(1)}$ is characterized as the most
general local functional belonging to the kernel of $S_0$ and 
to the orthogonal complement of ${\rm Im} \, S_0$.
Written on a basis $\{ {\cal C}_k \}_{k \geq 1}$ of
${\rm ker } \ S_0 \cap ({\rm Im } S_0)^\perp$, one gets
\begin{eqnarray}
Y^{(1)}=\sum_{k \geq 1} \lambda_k {\cal C}_k \ ,
\label{expy}
\end{eqnarray}
for some coefficients $\lambda_k$.
This expansion separates truly anomalous terms ($Y^{(1)}$) from
spurious ones ($S_0({\cal C}_0)$).
The latter can be canceled by a suitable redefinition of the
first-order counter-terms
entering in the construction of $\G^{(1)}$.

\subsection{The massive case}

In the massive HK model eq. (\ref{e2}) is modified as follows
\begin{eqnarray}
S_\G S(\G) = \int d^4x \, 
\left ( \square c + {ev \over \alpha} {\delta \G \over \delta J_2} \right )
{\delta \G \over \delta \bar c} \, .
\label{e9}
\end{eqnarray}
This identity is valid for any functional $\G$,
without restrictions as $S(\Gamma)=0$ or 
the ghost equation~(\ref{e8}).

Taking into account eq.(\ref{e8}) we get
\begin{eqnarray}
S_\G S(\G) = -{M^2 \over \alpha} \int d^4x \, c {\delta \G \over \delta \bar c} \, .
\label{e10}
\end{eqnarray}
The linearized ST operator $S_0$ is no more nilpotent; for any functional
$F$ satisfying the ghost equation (\ref{e8}) we have now
\begin{eqnarray}
S_0^2(F) = -{M^2 \over \alpha} \int d^4x \, c {\delta F \over \delta \bar c} \, .
\label{e11}
\end{eqnarray}
At the classical level the STI are satisfied:
\begin{eqnarray}
S(\G^{(0)}) = 0 \, .
\label{e12}
\end{eqnarray}
At the first order in perturbation theory eq.(\ref{e10}) gives, taking into account
eq.(\ref{e12}):
\begin{eqnarray}
S_0 ( S(\G)^{(1)} ) = -{M^2 \over \alpha} \int d^4x \, c {\delta \G^{(1)} \over \delta \bar c} \, .
\label{e13}
\end{eqnarray}
Noticing that
\begin{eqnarray}
S_0 ( S(\G)^{(1)} ) = S_0^2 (\G^{(1)}) \, ,
\label{e14}
\end{eqnarray}
we conclude that eq.(\ref{e11}) 
with $F=\G^{(1)}$ is embodied in eq.(\ref{e10}), under the assumption
of the ST invariance of the classical action in eq.(\ref{e12}).

If we adopt a regularization consistent with locality, power-counting and all other
unbroken symmetries of the theory (e.g. Lorentz invariance, C-parity, etc.), 
by using the Quantum Action Principle (QAP) \cite{QAP} we conclude that the R.H.S. of eq. (\ref{e13}) is zero.
Indeed, from the QAP and the power counting theorem we know that 
$\int d^4x \, c${\large ${\delta \G^{(1)} \over \delta \bar c}$} is a local C-even,
Lorentz invariant functional
and has dimension less or equal four and FP charge equal two. There are no terms with these properties,
so we get the following equation:
\begin{eqnarray}
S_0^2 (\G^{(1)}) = 0 \, ,
\label{e14ter}
\end{eqnarray}
i.e. even in the massive (non-nilpotent) case the QAP and the power-counting 
imply that the linearized ST operator $S_0$ is nilpotent on the space of 
the first-order quantum corrections $\G^{(1)}$.


Let us come back to the study of eq.(\ref{e13}). Since we know that its R.H.S. is zero,
thanks to the QAP and the power counting theorem, we have the following equation for
the breaking terms $ {\cal A}_1 = S(\G)^{(1)}$:
\begin{eqnarray}
S_0 ({\cal A}_1) = 0 \, .
\label{e14quat}
\end{eqnarray}
Since  even in the massive case $S_0$ is nilpotent on the action-like
functionals, one
can apply the same decomposition of eq.(\ref{e5}).
In the Abelian HK model there is just one cohomologically non trivial
insertion,
$\int d^4x \, \bar c c \partial_\mu c A^\mu$.  It has the right quantum numbers
and the correct exact symmetries of the theory (it is Lorentz-invariant and
C-even). However, it can be excluded thanks to the ghost equation (\ref{e8}),
the QAP and the power counting.
For $n \geq 1$ the ghost equation (\ref{e8}) can be written in the form
\begin{eqnarray}
{\cal G} \G^{(n)} = 0 \, .
\label{e14sexies}
\end{eqnarray}
Using  eq.(\ref{e14sexies}) and (\ref{e14octies}) we get:
\begin{eqnarray}
{\cal G} S_0(\G^{(1)}) = - S_0({\cal G} \G^{(1)}) = 0 \, .
\label{e14nonies}
\end{eqnarray}
By power counting $S_0(\G^{(1)})$ cannot contain the external source $J_2$ and
from eq. (\ref{e14nonies}) we conclude
\begin{eqnarray}
{\delta \over \delta \bar c} S_0(\G^{(1)}) = 0 \, .
\label{e14decies}
\end{eqnarray}
Thus the non-trivial breaking term $\int d^4x \, \bar c c \partial_\mu c A^\mu$ is not present
and the HK model turns out to be non-anomalous; suitable counter-terms can be constructed
at the first order in perturbation theory (actually at all orders), by which the STI can be restored (see \cite{FG},
\cite{FGQ}).

It is worthwhile noticing that on purely algebraic grounds there are no reasons to
exclude the anomalous insertion $\int d^4x \, \bar c c \partial_\mu c A^\mu$.
To this extent, the properties of the renormalization procedure, dictated by the QAP and the power
counting theorem, are essential.

Suppose now that the STI have been restored up to the $(n-1)$-th order in
perturbation theory, i.e.
we assume that
suitable counter-terms have been added iteratively to $\G^{(j)}$, $j=1, \dots, k$, 
in order to restore the STI till order $n-1$:~\footnote{Of course,
this is possible only in the absence of anomalies, as it is in the
Abelian HK model; in this case it can be proven that the restoration of the STI
to the $n$-order doesn't change the counter-terms needed to recover the STI
up to the $(n-1)$-th order, and can be performed, if the STI are fulfilled till
order $n-1$, by a proper choice of counter-terms at the $n$-order only.}
\begin{eqnarray}
S(\GG^{(k)}) = 0, ~~~ k =1,2, \dots, n-1
\label{e14bis}
\end{eqnarray}
$\GG^{(k)}$ denotes the correct symmetric effective
action at the $k$-th order in perturbation theory.

Then eq. (\ref{e10}) becomes
\begin{eqnarray}
S_0 ( S(\G)^{(n)} ) = -{M^2 \over \alpha} \int d^4x \, c {\delta \G^{(n)} \over \delta \bar c}
\label{e15}
\end{eqnarray}
or
\begin{eqnarray}
S_0 \left(
   S_0(\G^{(n)}) + \sum_{j=1}^{n-1} (\GG^{(n-j)}, \GG^{(j)} )
\right) = -{M^2 \over \alpha} \int d^4x \, c {\delta \G^{(n)} \over
   \delta \bar c}\ .
\label{e16}
\end{eqnarray}
Taking into account eq.(\ref{e11}) we arrive at the following 
consistency condition for the lower orders parts of the effective action:
\begin{eqnarray}
S_0 \left( \sum_{j=1}^{n-1} (\GG^{(n-j)}, \GG^{(j)} ) \right) = 0\ .
\label{e17}
\end{eqnarray}
This consistency condition is a consequence of the form of the linearized ST operator and
of the lower order requirements (\ref{e14bis}). Again, it relies on the use of the QAP
to ensure the fulfillment of STI at lower orders in perturbation
theory.

\section{Higher orders}

We now consider eq.(\ref{e10}) at the second order in perturbation
theory.  We do not
assume that the STI have been restored at the first order. Then
eq.(\ref{e10}) can be written as
\begin{eqnarray}
S_0 (S(\G)^{(2)}) + S_{\G^{(1)}} (S(\G)^{(1)}) =
 -{M^2 \over \alpha} \int d^4x \, c
{\delta \G^{(2)} \over \delta \bar c}\ .
\label{e18}
\end{eqnarray}
We show that, if we use a renormalization scheme where the QAP holds,
the R.H.S. of eq. (\ref{e18}) is zero.
We have to verify that
\begin{eqnarray}
-{M^2 \over \alpha} \int d^4x \, c {\delta \G^{(n)} \over \delta \bar
c} = 0\ ,
\label{v1}
\end{eqnarray}
for all $n$.
For $n=0$ the classical action $\G^{(0)}$ 
(appendix \ref{modello})
satisfies eq.(\ref{v1}), as it can be checked by explicit computation.
Suppose now that eq.(\ref{v1}) is verified till order $n-1$:
\begin{eqnarray}
-{M^2 \over \alpha} \int d^4x \, c {\delta \G^{(k)} \over \delta \bar c} = 0,
~~~~~ k=1,\dots, n-1
\label{v2}
\end{eqnarray}
By using the QAP, at the next order in perturbation theory we get:
\begin{eqnarray}
-{M^2 \over \alpha} \int d^4x \, c {\delta \G^{(n)} \over \delta \bar c} = 
\int d^4x \, \Delta(x), 
\label{v3}
\end{eqnarray}
where $\int d^4x \, \Delta(x)$ is an integrated Lorentz invariant
local polynomial $\Delta(x)$ in the fields of the theory. $\Delta(x)$ has
dimension $\leq 4$, $FP$-charge $+2$ and it obeys 
all the exact symmetries of the model.
Since there are no terms with these properties ($\int d^4x \, c c = 0$,
$\int d^4x \, c \square c =0$, $\int d^4x A_\mu c \partial^\mu c$ is
excluded by $C$-parity, and so on), we conclude that at the $n$-th order
\begin{eqnarray}
-{M^2 \over \alpha} \int d^4x \, c {\delta \G^{(n)} \over \delta \bar c} = 0.
\label{v4}
\end{eqnarray}
This in turn implies that eq.(\ref{v1}) holds true for all $n$.

This result can be demonstrated by a direct analysis of the
Feynman graphs, arising in the perturbative expansion of $\G$.

Moreover, the QAP also implies that $S(\G)^{(1)}$ cannot depend
on external sources. 
Thus eq.(\ref{e18}) simplifies to
\begin{eqnarray}
S_0 (S(\G)^{(2)}) = - \sum_i \int d^4x \, \frac{\delta \G^{(1)}}{\delta J_i(x)}
 {\delta (S(\G)^{(1)}) \over \delta \phi_i(x)} 
\label{e19}\ ,
\end{eqnarray}
where the sum is extended to all the fields $\phi_i$ whose BRST variation is non linear.

From eq.(\ref{e19}) we see that, 
if the STI have been restored at the first order (i.e. $S(\G)^{(1)}=0$), 
the Wess-Zumino consistency condition holding for $S(\G)^{(1)}$
is true for $S(\G)^{(2)}$ too.
In particular, $S(\G)^{(2)}$ is also local, like $S(\G)^{(1)}$.

On the contrary, if $S(\G)^{(1)} \not =0$, 
the Wess-Zumino consistency condition for $S(\G)^{(2)}$ 
is modified by the R.H.S. of eq.(\ref{e19}), which now is non-zero.
Moreover, we show in eq.(\ref{f15}) that 
eq.(\ref{e19}) implies that $S(\G)^{(2)}$ 
receives non-local contributions, arising from
the insertion of the local functional
$\int d^4x$ {\large${\delta (S(\G)^{(1)}) \over \delta \phi_i(x)}$} on
the non-local quantities {\large ${\delta \G^{(1)} \over {\delta J_i(x)}}$}. 

In order to simplify the notations,  we define $X \equiv S(\G)^{(2)}$. 
We also define (in the momentum space) the fourth-order
differential operator 
\begin{eqnarray}
{\cal P} (\cdot) \equiv {\delta^4 (\cdot) \over  \delta \bar c(q) \delta c(s)
 c(t) c(r)}
\label{f12}
\end{eqnarray}
We apply ${\cal P}$ on both sides of eq.(\ref{e19}) and then set the fields
(including external sources) to zero. By using $C$-parity and the fact
that the $FP$-charge of $\G$ is zero, we get~\footnote{We use a short
hand for functional derivatives:
$Y_{c(p)}$ denotes $\frac{\delta Y}{\delta c(p)}$, 
$Y_{c(p)\phi_1(s)}$ denotes $\frac{\delta^2 Y}{\delta c(p)\delta\phi_1(s)}$,
and so on.} 
\begin{eqnarray}
%
%
\left. {\cal P} S_0(X) \right |_{\varphi =0} & = &
\left . - i r_\mu X_{\bar c(q) c(s) c(t) A_\mu(r)}\right|_{\varphi =0}
\left . + ev X_{\bar c(q) c(s) c(t) \phi_2(-r)}\right|_{\varphi =0}
\left . + e^2v X_{c(s)c(t) J_1(-q-r)}\right|_{\varphi =0}
\nonumber\\
& & + {\rm cycl.~permutations~ of~(s,t,r)}
\label{f13}
\end{eqnarray}
and
\begin{eqnarray}
&&  \left . {\cal P} \left [ -\sum_i  
\left ( \int {d^4p \over (2\pi)^4} 
\, \Gamma^{(1)}_{J_i(p)} 
 S(\G)^{(1)}_{\phi_i(p)} \right ) \right ] \right |_{\varphi = 0} =  \nonumber \\
&& ~~~~~~ \int {d^4p \over (2\pi)^4} \, 
\Big ( \left.
\Gamma^{(1)}_{\bar c(q) c(r)c(t)J_2(p)}\right|_{\varphi = 0}
\left. S(\G)^{(1)}_{c(s)\phi_2(p)}\right|_{\varphi = 0} +
{\rm cycl.~permutations~of~(s,t,r)}
\Big )
\label{f6}
\end{eqnarray}
Taking into account the conservation of momenta in 1PI Green functions,
we see that the R.H.S. of eq.(\ref{f6})
is not zero for $-s+q+r+t =0$, for
$-t+q+r+s=0$, or for $-r+s+q+t=0$.

%
%
%

\begin{figure}
\epsfxsize=90mm
\centerline{\epsffile{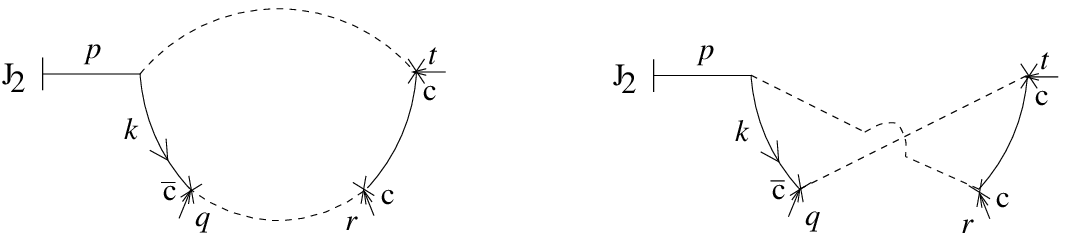}}
\caption{Graphs contributing to 
$\left . \G^{(1)}_{\bar c(q) c(r) c(t) J_2(p)} \right |_{\varphi=0}$ 
- Type I}\label{figura3}
\end{figure}

\begin{figure}
\epsfxsize=90mm
\centerline{\epsffile{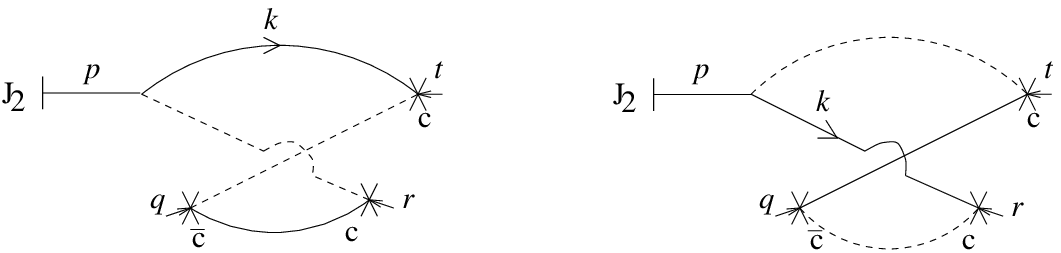}}
\caption{Graphs contributing to 
$\left . \G^{(1)}_{\bar c(q) c(r) c(t) J_2(p)} \right |_{\varphi=0}$  - Type II}\label{figura4}
\end{figure}

The amplitude $\left.\G^{(1)}_{\bar c(q) c(t) c(r) J_2(p)}\right|_{\varphi=0}$
gets contributions from the graphs in Figures \ref{figura3}
and \ref{figura4}. All external momenta
are understood to be incoming. The relevant Feynman rules are briefly discussed
in appendix~ \ref{feyn}.
%
%
%
Working out the explicit form of
$\left.\G^{(1)}_{\bar c(q) c(t) c(r) J_2(p)}\right|_{\varphi=0}$
we have (after performing the Wick rotation):
\begin{eqnarray}
\left.\G^{(1)}_{\bar c(q) c(t) c(r) J_2(p)}\right|_{\varphi=0}
 & = & i {e^7 v^3 \over \alpha^2} 
\int d^4 k \, \Big ( 
{1 \over k^2 +m_1^2} {1 \over (k+q)^2 +m_g^2} 
{1 \over (k+q+r)^2 + m_1^2} {1 \over (k-p)^2 +m_g^2} 
\nonumber \\
& & ~~~~~~~~~~~~~~~~~~~~ + {1 \over k^2 +m_1^2} {1 \over (k+r)^2 +m_g^2} 
{1 \over (k+q+r)^2 + m_1^2} {1 \over (k-p)^2 +m_g^2} 
\nonumber \\
& & ~~~~~~~~~~~~~~~~~~~~ - (r \leftrightarrow t) \Big )
\cdot \delta^{(4)}(p+q+r+t)
\label{e20}
\end{eqnarray}
All integrals are convergent (no subtraction required). For general momenta
$p,q,r,t$ the R.H.S. of eq.(\ref{e20}) is non-zero.
Moreover, it is non-polynomial
in the independent external momenta: by applying Weinberg's theorem \cite{wein} 
we conclude that for non-exceptional external momenta the amplitude
$\left.\G^{(1)}_{ \bar c(Qq) c(Qt) c(Qr) J_2(Qp)}\right|_{\varphi=0}$
  behaves as $\sim Q^{-4}$ for
$Q \rightarrow \infty$ and fixed $p,q,r,t$ (in the Euclidean region).

A direct calculation of $S(\G)^{(1)}$, obtained by applying the method
described in \cite{FGQ}, shows that 
\begin{eqnarray}
\left. S(\G)^{(1)}_{c(s)\phi_2(p)}\right|_{\varphi=0} =
 \left(a + b s^2\right)\delta^{(4)}(s+p)
\label{f14}
\end{eqnarray}
for some $c$-numbers $a,b$ (depending on the intermediate renormalization
scheme used).

By using eqs.(\ref{f13}), (\ref{f6}) and (\ref{f14}) we obtain the following
equation
\begin{eqnarray}
&&\left . - i r_\mu X_{\bar c(q) c(s) c(t) A_\mu(r)}\right|_{\varphi =0}
+\left .  ev X_{\bar c(q) c(s) c(t) \phi_2(-r)}\right|_{\varphi =0}
\left . + e^2v X_{c(s)c(t) J_1(-q-r)}\right|_{\varphi =0}
\nonumber\\
&&~~~~~~~ +{\rm cycl.~permutations~of~(s,t,r)} \nonumber \\
&&~~~~~~~ =
 (a + bs^2)
\Gamma^{(1)}_{\bar c(q) c(r)c(t)J_2(-s)}
+{\rm cycl.~permutations~of~(s,t,r)}
\label{f15}
\end{eqnarray}
Eq.(\ref{e20}) implies that the R.H.S. of eq.(\ref{f15}) is 
 non-polynomial in the
variables $s,q,r,t$. Thus at least one of the amplitudes
$X_{\bar c c c A_\mu}, X_{\bar c c c \phi_2},
X_{cc J_1}$ is non-polynomial in the external momenta $s,q,r,t$. 
This in turn implies that
$S(\G)^{(2)}$ is non-local.

From eq.(\ref{f15}) we learn that some of the terms arising in the
expansion of $S(\G)^{(2)}$ on a basis of the fields $\Phi$ 
and the external sources $\Phi^*$
of the theory must contain an arbitrarily high number of derivatives.
We will show that the expansion of $S(\G)^{(2)}$ on a basis of
$\Phi, \Phi^*$  must also contain terms with an arbitrarily
high number of fields $\phi_1$.

\begin{figure}
\epsfxsize=90mm
\centerline{\epsffile{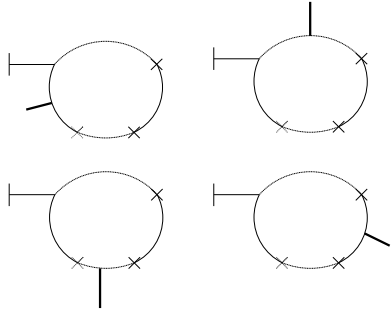}}
\caption{Graphs generated by insertion of $\phi_1(w_1)$}\label{figura5}
\end{figure}

Consider the insertion of $k$ fields $\phi_1$ along
the $\bar c c$-lines or the $\phi_1 \phi_1$-lines in the graphs
shown in figures \ref{figura3} and \ref{figura4}.
It is worthwhile performing the construction of these insertions in a recursive
way. 
It is possible to insert a leg $\phi_1(w_1)$ (carrying momentum $w_1$)
in the graph on the left of figure \ref{figura3}
by cutting a $\bar c c$-propagator or by cutting a $\phi_1 \phi_1$-line.
The graph on the left of figure \ref{figura3} thus generates four graphs
contributing to the 1-PI amplitude $\left.\G^{(1)}_{\bar c(q) c(t) c(r) J_2(p) \phi_1(w_1)}\right|_{\varphi=0}$.
They are shown in figure \ref{figura5}.
%
%
%
The same construction can be applied to all other graphs appearing in figures \ref{figura3}
and \ref{figura4}, yielding a family ${\cal E}^1$ of graphs contributing
to  $\left.\G^{(1)}_{\bar c(q) c(t) c(r) J_2(p) \phi_1(w_1)}\right|_{\varphi=0}$.
Moreover, the graphs in ${\cal E}^1$ exhaust all possible graphs
contributing to $\left.\G^{(1)}_{\bar c(q) c(t) c(r) J_2(p) \phi_1(w_1)}\right|_{\varphi=0}$.
Indeed, if ${\cal T}$ is a graph contributing to $\left.\G^{(1)}_{\bar c(q) c(t) c(r) J_2(p) \phi_1(w_1)}\right|_{\varphi=0}$, the leg $\phi_1(w_1)$ must be inserted either on a $\phi_1\phi_1$-line
or on a $\bar c c$-line. Thus removing the insertion
of $\phi_1(w_1)$ and gluing together the two propagators $\phi_1\phi_1$ -
$\phi_1 \phi_1$ or $\bar c c$-$\bar c c$ of the line
with the $\phi_1(w_1)$ insertion into a single $\phi_1 \phi_1$
or $\bar c c$ propagator respectively yields a graph contributing to
$\left.\G^{(1)}_{\bar c(q) c(t) c(r) J_2(p)}\right|_{\varphi=0}$.
 The latter must be one of the
graphs in figures \ref{figura3} or \ref{figura4}. Hence ${\cal T}$
belongs to ${\cal E}^1$.

From an analytic point of view, the construction of graphs
in ${\cal E}^1$ amounts to the replacement
\begin{eqnarray}
{1 \over k^2 +m_g^2} \rightarrow {1 \over {(k+w_1)^2+m_g^2}}
{1 \over k^2 +m_g^2}
\label{f30}
\end{eqnarray}
(for a $\phi_1(w_1)$-insertion on a $\bar c c$-line) or
\begin{eqnarray}
{1 \over k^2 +m_1^2} \rightarrow {1 \over {(k+w_1)^2+m_1^2}}
{1 \over k^2+m_1^2}
\label{f31}
\end{eqnarray}
(for a $\phi_1(w_1)$-insertion on a $\phi_1 \phi_1$-line).

No new cancellations arise from these replacements, as it can be seen
from the explicit computation of the associated integrals, leading to
a straightforward generalization of eq.(\ref{e20}).
So one can conclude that $\left.\G^{(1)}_{\bar c(q) c(t) c(r) J_2(p) \phi_1(w_1)}\right|_{\varphi=0}$
is non-zero.

This construction can be applied to the graphs of ${\cal E}^1$
to generate the family ${\cal E}^2$, whose elements are graphs contributing to 
$\left.
\G^{(1)}_{J_2(p) \bar c(q) c(r) c(s) \phi_1(w_1) \phi_1(w_2)}
\right|_{\varphi=0}$.
Again, the same remarks as before apply and one finds out that
$\left.
\G^{(1)}_{J_2(p) \bar c(q) c(r) c(s) \phi_1(w_1) \phi_1(w_2)}
\right|_{\varphi=0} \not = 0
$.
The recursion can be iterated till the $k$ insertions of $\phi_1(w_j)$,
$j=1,\dots,k$
have been completed.

We now introduce an extension of the operator ${\cal P}$ defined
in eq.(\ref{f12}):
\begin{eqnarray}
{\cal P}_k (\cdot) \equiv 
{\delta^{4+k} (\cdot) \over \delta \phi_1(w_k) \delta \phi_1(w_{k-1}) \dots
\delta \phi_1(w_1) \delta \bar c(q) \delta c(s) \delta c(t) \delta c(r)}.
\label{f32}
\end{eqnarray}
By applying ${\cal P}_k$ to eq.(\ref{e19}) and setting next the fields
to zero, one gets the following set of equations:
\begin{eqnarray}
{\cal P}_k \left [ S_0 (S(\G))^{(2)} \right ]_{\varphi=0} = - \sum_i \int d^4x \, {\cal P}_k \left [ {\delta \G^{(1)} \over \delta J_i(x)} {\delta (S(\G)^{(1)}) \over \delta \phi_i(x)} \right]_{\varphi=0}
\label{f33}
\end{eqnarray}
For $k=0$ we recover eq.(\ref{f15}). The R.H.S. of eq.(\ref{f33}) contains
functions of the form
\begin{eqnarray}
&& S(\G)^{(1)}_{c(p_1)\phi_2(p_2)\prod_{a=1}^k \phi_1(q_a)} = 
Q(p_1,p_2,q_a) \delta^{(4)}(p_1+p_2+\sum_{a=1}^k q_a),
\label{f34}
\end{eqnarray}
where $Q(p_1,p_2,q_a)$ is a polynomial of degree at most $2$
in $p_1,p_2,q_a$.
These functions are zero for $k > 4$, as it can be seen by a direct
calculation applying the method in \cite{FGQ}.

The R.H.S. of eq.(\ref{f33}) has the same feature as the
R.H.S. of eq.(\ref{f15}): each configuration of external momenta,
compatible with the delta functions contained in the R.H.S. of
eq.(\ref{f33}), picks out one and only one of the terms which arise
in the expansion of the R.H.S. of eq.(\ref{f33}) in terms of the amplitudes
$\left.
\G^{(1)}_{c(p_1) \bar c(p_2) c(p_3) J_2(p_4) \prod_{a=1}^j \phi_1(q_a)}
\right|_{\varphi=0}$,
$j=1,\dots,k$.
Thus, having shown that 
\begin{eqnarray}
\left.
\G^{(1)}_{c(p_1) \bar c(p_2) c(p_3) J_2(p_4) \prod_{a=1}^j \phi_1(q_a)}
\right|_{\varphi=0} \not = 0,
\label{y2}
\end{eqnarray}
we conclude that the R.H.S. of
eq.(\ref{f33}) is non-zero for every $k$.
Eq.(\ref{f33}) implies that,
in the expansion of $S(\G)^{(2)}$ on a basis of $\Phi,\Phi^*$,
there are non-zero terms associated with monomials
containing an arbitrary number of $\phi_1$ fields.

This has some  
interesting consequences.
We have shown that, if improper finite counter-terms in $\G^{(1)}$ are chosen, 
at the second
order in perturbation theory the STI
\begin{eqnarray}
S(\G)^{(2)} = {\cal A}_2 
\label{y1}
\end{eqnarray}
are broken by a non-local functional ${\cal A}_2$.
Eq. (\ref{y1})
gives rise upon differentiation with respect to a set of fields
and external sources  $\{\Phi^I,\Phi^*_I\}_{I \in {\cal I}}$ (with $I$
running in the set of indices ${\cal I}$) to a number of relations
among 1-PI Green functions, once we set $\Phi^I = \Phi^*_I =0$ after
taking the relevant derivatives.

We can expand ${\cal A}_2$ on a basis of monomials in $\Phi,\Phi^*$
and their derivatives.
Since ${\cal A}_2$ is non-local, an infinite number of monomials
appears in this expansion. It may happen that there is a maximum finite
number $O$ of $\Phi,\Phi^*$, appearing in every monomial of the expansion.
In this case, the expansion is infinite because 
it contains monomials
with arbitrarily high order derivatives.
Thus we can differentiate eq.(\ref{y1}) with respect to a number
of fields greater than $O$, yielding the same result as if
$S(\G)^{(2)}=0$.
Only a finite number of relations among 1-PI Green functions,
valid in the invariant case $S(\G)^{(2)}=0$, is altered
by this type of non-local breaking terms ${\cal A}_2$. 
Notice that this is the same behavior one
has when the breaking is local.

It may also happen that in the expansion of ${\cal A}_2$ 
there appear monomials with an arbitrarily high number of $\Phi,\Phi^*$.
Now an infinite set of relations among 1-PI Green functions,
derived from eq.(\ref{y1}) upon differentiation, is changed with
respect to the invariant case.
In this sense, violation of locality by arbitrarily high number of 
$\Phi,\Phi^*$
is more severe than violation of locality by arbitrarily high number of
derivatives {\em only}.

We briefly comment on the results of this section. 
Had we restored the STI at the first order in perturbation theory, 
eq.(\ref{e18}) would have read
\begin{eqnarray}
S_0(S(\G)^{(2)})=0.
\label{f34bis}
\end{eqnarray}
If $S(\G)^{(1)}=0$, the same Wess-Zumino consistency condition holds true
both for $S(\G)^{(1)}$ (see eq.(\ref{e4})) and $S(\G)^{(2)}$.
In particular, $S(\G)^{(2)}$ is local.
Moreover, eq.(\ref{f15}) shows that if
$S(\G)^{(1)} \not =0$, $S(\G)^{(2)}$ receives non-local contributions.
Thus a necessary and sufficient condition for $S(\G)^{(2)}$ to be local
is that $S(\G)^{(1)}=0$.
Notice that one can impose in the Abelian HK model
$S(\G)^{(1)}=0$ because the model does not possess physical anomalies.
 
This result admits a wider generalization.
Suppose that the gauge theory under investigation is truly anomalous.
Then at the second order in perturbation theory the consistency
condition obeyed by $S(\G)^{(2)}$ is eq.(\ref{e19}), where now
$S(\G)^{(1)}$
is non zero for any choice of the first order action-like counterterms.
An argument similar to the one leading to eq.(\ref{f13}) thus entails
that $S(\G)^{(2)}$ must be non local, because of the contributions coming
from
$S_{\G^{(1)}}(S(\G)^{(1)})$.~\footnote{
In \cite{white} it was argued that, if one
recovers the spurious contributions to the first order anomaly,
$S(\G)^{(2)}$ can be made local and chosen in such a way that it satisfies
the same Wess-Zumino
consistency condition as $S(\G)^{(1)}$.
To match these requirements, one needs to introduce 
the first order truly anomalous (local) terms as interaction vertices 
in the quantum effective
action $\G^{(1)}$, by coupling them to external sources of
{\em negative} dimension.

In our opinion, this procedure generates a set of Feynman rules
which spoil the validity of the QAP at the next order. In particular,
at the next order $S_0(S(\G)^{(2)})$ gets non-local
contributions, which in our framework
are embodied in $S_{\G^{(1)}}(\G^{(1)})$.

Notice that, if the first order physical anomalies
had dimension $\leq 4$ (which is not forbidden by the QAP, saying only
that they must have dimension less or equal to  $5$),
no troubles  would arise in including them as vertices
in $\G^{(1)}$. They could be coupled to external sources
with non-negative dimension.
Actually, 
truly anomalous terms have dimension $5$ only
(at least for a gauge group without Abelian factors) \cite{anomalies}.
This in turn implies that they cannot just be thought as 
new interaction vertices, since these vertices
must contain external sources with dimension
$-1$.}
%



\section{Conclusions}

In this paper  we have shown that, if the action-like counter-terms
entering in $\G^{(1)}$ are not properly chosen, even a physically
non-anomalous theory exhibits a non-local second order anomaly.
This anomaly cannot be restored by local second order counterterms.
Thus, an improper choice of the finite part of the first order
counter-terms renders a first-order physically non-anomalous theory
a second order truly anomalous one.

Moreover, we have argued that, if one starts with a truly anomalous theory,
locality of the STI breaking terms is satisfied at the first order
in perturbation theory only, no matter which renormalization scheme
is adopted. 


We conclude that locality of the STI breaking terms can be
maintained to all orders if and only if there are no truly anomalous
terms at the first order in perturbation theory.
  
Finally, we have shown that strict nilpotency of the BRST transformations
(and consequently of the linearized ST operator) is not an essential
requirement in order to perform the characterization of the
STI breaking terms, independently on the order of the perturbative
expansion.

\vskip 0.5 cm

{\Large\bf Acknowledgements}

We acknowledge Professor ~R.~Ferrari for useful
discussions.
%

\vfill\eject
\appendix

\section{Classical action} \label{modello}
The classical action for the HK model in the on-shell formalism is 
\begin{eqnarray}
&&
\Gamma^{(0)} = \int d^4x \Big [- {1\over 4}F_{\mu\nu}^2
+ {{e^2v^2}\over 2} A_\mu^2
\nonumber\\
&&
-{\alpha \over 2} \partial A^2 +  \alpha {\bar c}\square c 
+ e^2v^2{\bar c}c + e^2 v {\bar c}c\phi_1
\nonumber\\
&&
+{1\over 2}((\partial_\mu\phi_1)^2 + (\partial_\mu\phi_2)^2)
-\lambda v^2 \phi_1^2 - 
{{e^2v^2}\over {2\alpha}} \phi_2^2
\nonumber\\
&&
+eA_\mu (\phi_2\partial^\mu \phi_1-\partial^\mu\phi_2\phi_1)
+ e^2v\phi_1 A^2 +{{e^2}\over 2}(\phi_1^2+\phi_2^2) A^2 
\nonumber\\
&&
- \lambda v \phi_1(\phi_1^2+\phi_2^2)
-{\lambda\over 4}(\phi_1^2+\phi_2^2)^2
\nonumber\\
&&
+ {\bar\psi}i\not \!\!\partial\psi +Gv{\bar\psi}\psi
+{e\over 2}{\bar\psi}\gamma_\mu\gamma_5\psi A^\mu
\nonumber\\
&&
+G{\bar\psi}\psi\phi_1
-iG{\bar\psi}\gamma_5\psi\phi_2
\nonumber\\
&&
+J_1 [-ec\phi_2] + J_2 ec(\phi_1+v)
+i{e\over 2}{\bar\eta}\gamma_5\psi c
+i{e\over 2}c{\bar\psi}\gamma_5 \eta
\nonumber \\
&& + {M^2 \over 2} A_\mu^2 + M^2 \bar c c 
- {M^2 \over 2 \alpha} (\phi_1^2 + \phi_2^2) \Big ] 
\end{eqnarray}

The explicit mass term is in evidence in the last line.

\vskip 0.5 cm

BRST transformations

\vskip 0.1 cm
Off-shell formalism
\begin{eqnarray}\label{BRST-off}
&&
s A_\mu = \partial_\mu c, ~~~ 
s \phi_1 = -ec \phi_2, ~~~ 
s \phi_2 = ec (\phi_1 +v) 
\nonumber \\
&&
s \psi = -i {e \over 2} \gamma_5 \psi c, ~~~ 
s \bar \psi = i {e \over 2} c \bar \psi \gamma_5, ~~~ 
s \bar c = B, ~~~ 
s B = -{M^2 \over \alpha} c, ~~~
s c = 0
\end{eqnarray}
In the on-shell formalism the $B$ field disappears and the BRST transformation of $\bar c$
becomes
\begin{eqnarray}\label{BRST}
s \bar c = \partial A + {ev \over \alpha} \phi_2, ~~~ 
\end{eqnarray}

\section{Feynman rules} \label{feyn}

We only recall the Feynman rules needed to evaluate the graphs
in Figures \ref{figura3} and \ref{figura4}.

Propagator for $\phi_1 \phi_1$
\begin{eqnarray}
\Delta_{\phi_1\phi_1}(p) = {i \over p^2 - m_1^2+i\epsilon},
~~~~~~ m_1^2 = 2\lambda v^2 + {M^2 \over \alpha}
\label{ae4}
\end{eqnarray}

The propagator for $\phi_1 \phi_1$ is denoted by a solid line.

\par
Propagator for $c \bar c$
\begin{eqnarray}
\Delta_{c \bar c}(p) = {-i \over \alpha \left ( p^2 - m_g^2 +i\epsilon \right)},
~~~~~~ m_g^2 = {e^2v^2 + M^2 \over \alpha}
\label{ae5}
\end{eqnarray}

The propagator for $c \bar c$ is denoted by a dashed line.

\vfill\eject
Vertices
\begin{figure}
\epsfxsize=50mm
\centerline{\epsffile{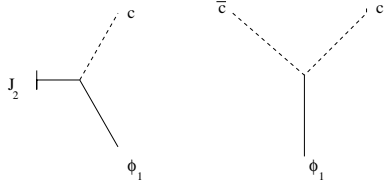}}
\caption{Vertices $eJ_2c\phi_1$ and $e^2v \bar c c \phi_1$}
\label{figura1}
\end{figure}
%
%
\begin{eqnarray}
ie (2\pi)^4 \delta^{(4)}(\mbox{incoming momenta})
\label{ae6}
\end{eqnarray}
for the vertex $e J_2 c \phi_1$ and
\begin{eqnarray}
ie^2v (2\pi)^4 \delta^{(4)}(\mbox{incoming momenta})
\label{ae7}
\end{eqnarray}
for the vertex $e^2v \bar c c \phi_1$ (see Figure \ref{figura1}).

\end{document}